# Rotation, Equivalence Principle, and GP-B Experiment

Wei-Tou Ni[1,2]

[1]*Shanghai United Center for Astrophysics (SUCA),*
*Shanghai Normal University, Shanghai 200234, China*
[2]*Center for Gravitation and Cosmology, Department of Physics,*
*National Tsing Hua University, Hsinchu, Taiwan, 30013, ROC*

**Abstract**
The ultra-precise Gravity Probe B experiment measured the frame-dragging effect and geodetic precession on four quartz gyros. We use this result to test WEP II (Weak Equivalence Principle II) which includes rotation in the universal free-fall motion. The free-fall Eötvös parameter $\eta$ for rotating body is $\leq 10^{-11}$ with four-order improvement over previous results. The anomalous torque per unit angular momentum parameter $\lambda$ is constrained to $(-0.05 \pm 3.67) \times 10^{-15}$ s$^{-1}$, $(0.24 \pm 0.98) \times 10^{-15}$ s$^{-1}$, and $(0 \pm 3.6) \times 10^{-13}$ s$^{-1}$ respectively in the directions of geodetic effect, frame-dragging effect and angular momentum axis; the dimensionless frequency-dependence parameter $\kappa$ is constrained to $(1.75 \pm 4.96) \times 10^{-17}$, $(1.80 \pm 1.34) \times 10^{-17}$, and $(0 \pm 3) \times 10^{-14}$ respectively.

Equivalence principles [1, 2, 3] are cornerstones in the foundation of gravitation theories. Galilei Equivalence Principle states that test bodies with the same initial position and initial velocity fall in the same way in a gravitational field. This principle is also called Universality of Free Fall (UFF) or Weak Equivalence Principle (WEP).

Since a macroscopic test body has 3 translational and 3 rotational degrees of freedom, true equivalence must address to all six degrees of freedom. We propose a second Weak Equivalence Principle (WEP II) to be tested by experiments. WEP II states that the motion of all six degrees of freedom of a macroscopic test body must be the same for all test bodies [4, 5]. There are two different scenarios that WEP II would be violated: (i) the translational motion is affected by the rotational state; (ii) the rotational state changes with angular momentum (rotational direction/speed) or species.

In the latter part of 1980's and early 1990's, a focus is on whether the rotation state would affect the trajectory. In 1989, Hayasaka and Takeuki [6] reported their results that, in weighing gyros, gyros with spin vector pointing downward reduced weight proportional to their rotational speed while gyros with spin vector pointing upward did not change weight. This would be a violation of WEP II if confirmed. Since the change in weight $\delta m$ is proportional to the angular momentum in this experiment, the violation could be characterized by the parameter ν defined to be $\delta m/I$ where $I$ is the angular momentum of the gyro. Soon after, Faller et al. [7], Quinn and Picard [8], Nitshke and Wilmarth [9], and Imanish et al. [10] performed careful weighing experiments on gyros with improved precision, but found only null results which are in disagreement with the report of Hayasaka and Takeuchi [6]. In 2002, Luo et al. [11] and Zhou et al. [12] set up interferometric free-fall experiments and found null results in disagreement with [6] also.

Table I compiles the experimental results. In the second and third columns, we list the parameter $\nu$ and the Eötvös parameter $\eta$ measured in each experiment. The Eötvös parameter $\eta$ is defined as $\delta m/m$. The angular momentum $I$ is given by $I = 2\pi f\, m\, r_{gyration}^2$ where $r_{gyration}$ (= [moment of inertia/$m$]$^{1/2}$) is the radius of gyration for the rotating body. Hence, we have the relation

$$\nu = \eta / (2\pi f\, r_{gyration}^2), \qquad (1)$$

where $f$ is the frequency of rotation of the gyro.

Table I. Test of WEP II regarding to trajectory using bodies with different angular momentum.

| Experiment | $\nu$ [s/cm$^2$] | $\lvert\eta\rvert$ | Method |
|---|---|---|---|
| Hayasaka-Takeuchi (1989) [6] | $(-9.8\pm0.9)\times10^{-9}$ for spin up, $\pm0.5\times10^{-9}$ for spin down | up to $6.8\times10^{-5}$ | weighing |
| Faller *et al* (1990) [7] | $\pm4.9\times10^{-10}$ | $< 9\times10^{-7}$ | weighing |
| Quinn-Picard (1990) [8] | $\lvert\nu\rvert \le 1.3\times10^{-10}$ | $< 2\times10^{-7}$ | weighing |
| Nitschke-Wilmarth (1990) [9] | $\lvert\nu\rvert \le 1.3\times10^{-10}$ | $< 5\times10^{-7}$ | weighing |
| Imanishi *et al.* (1991) [10] | $\lvert\nu\rvert \le 5.8\times10^{-10}$ | $< 2.5\times10^{-6}$ | weighing |
| Luo *et al.* (2002) [11] | $\lvert\nu\rvert \le 3.3\times10^{-10}$ | $\le 2\times10^{-6}$ | free-fall |
| Zhou *et al.* (2002) [12] | $\lvert\nu\rvert \le 2.7\times10^{-11}$ | $\le 1.6\times10^{-7}$ | free-fall |
| Everitt *et al.* (2011) [13] | $\lvert\nu\rvert \le 6.6\times10^{-15}$ | $\le 1\times10^{-11}$ | free-fall |

For rotating bodies, Gravity Probe B (GP-B) experiment [13-19] has the best accuracy. GP-B, a space experiment launched 20 April 2004, with 31 years of research and development, 10 years of flight preparation, a 1.5 year flight mission and 5 years of data analysis, has arrived at the final experimental results for this landmark testing two fundamental predictions of Einstein's theory of General Relativity (GR), the geodetic and frame-dragging effects, by means of cryogenic gyroscopes in Earth orbit. The spacecraft carries 4 gyroscopes (quartz balls) pointing to the guide star IM Pegasi in a polar orbit of height 642 km. GP-B was conceived as a controlled physics experiment having mas/yr stability ($10^6$ times better than the best modeled navigation gyroscopes) with numerous built-in checks and methods of treating systematics.

With GP-B accuracy, the impact of its implication on the tests of various physics will need some time to investigate. Here, we use the GP-B results to test WEP II and to constrain relevant parameters. The results of the experiment are compiled in Table II [13, 20, 21]. The quartz gyroscope has a diameter of 3.81 cm. The rotation (spin) rates of four gyros are tabulated in the second column of Table II. The four quartz gyros were initially aligned to the bore sight of the telescope pointing to the guide star IM Pegasi (HR 8703). Gyroscopes 1 and 3 had their spin axes (using the right hand rule) pointed toward the guide star (positive spin rate) while Gyroscopes 2 and 4 had their spin axes pointed in the opposite direction from the direction to the guide star (negative spin rates). The spin-down rates of the four gyroscopes are tabulated in the third column of Table II. In calculating the $\nu$ and $\eta$ parameters for GP-B, we use the data listed in the Gravity Probe B Quick Facts [22]. There are four gyroscopes (G1, G2, G3 and G4) with one of them also as a drag-free test body. The drag-free performance is better than $10^{-11}$ g. In a more detailed analysis, the relative acceleration of different gyros with different speed needs to be deduced from levitating feedback data and local space gravity distribution. With this analysis, the results for relevant frequencies could be better. Here we take $10^{-11}$ g as an upper bound of the Eötvös parameter $\eta$. With its precision, GP-B gives a constraint on $v$ much better than previous experiments on earth. The result of GP-B is about 4 orders better than the second best experimental result for rotating bodies (Table I).

Table II. Results of GP-B experiment

| Gyro | $f_s$ (Hz) | d$f_s$/dt (μHz/hr) | $r_{NS}$ (mas/yr) | $r_{WE}$ (mas/yr) |
|---|---|---|---|---|
| G1 | 79.39 | -0.57 | -6588.6±31.7 | -41.3±24,6 |
| G2 | -61.82 | -0.52 | -6707.0±64.1 | -16.1±29.7 |
| G3 | 82.09 | -1.30 | -6610.5±43.2 | -25.0±12.1 |
| G4 | -64,85 | -0.28 | -6588.7±33.2 | -49.3±11.4 |

To test WEP II regarding to the rotational state changes with different angular momentum (rotational direction/speed) or species, one needs to measure the rotational direction and speed very precisely with respect to time. GP-B has four gyros rotating with different speeds and has measured the rotational directions very precisely. The quartz rotors have been placed in high-vacuum housing with a very long spin-down time. The WEP II violation parameter λ for a test body is defined to be the anomalous torque $\Gamma_a$ on the rotating body divided by its angular momentum Iω:

$$\lambda = \Gamma_a / (I\omega). \tag{2}$$

Anomalous torque is equal to anomalous angular momentum change divided by time:

$$\Gamma_a = d(I\omega)/dt. \tag{3}$$

Angular momentum change divided by angular momentum and time gives anomalous angular drift in the transverse (to rotation axis) direction, while gives anomalous rate of change of the rotation speed in the axial direction.

For the anomalous torque, we use a simple phenomenological model which assumes linear dependence (with parameters $\kappa$ and $\lambda$) in the rotational speed. With this assumption, the anomalous torque contributions to the drifts are

$$r_{NS}^A = \kappa_{NS} f_s + \lambda_{NS},$$
$$r_{WE}^A = \kappa_{WE} f_s + \lambda_{WE}. \tag{4}$$

In this model, the weighted fitting to $r_{NS}$ and $r_{WE}$ as functions of $f_s$ should include the anomalous terms. Two fitting results are shown in Figure 1 and Figure 2. From fitting, the parameters are determined to be

$$\kappa_{NS} = -0.1142 \pm 0.3235 \text{ mas yr}^{-1} \text{ Hz}^{-1} = (1.75 \pm 4.96) \times 10^{-17},$$
$$\lambda_{NS} + r_{NS}^{GR} = -6605.8 \pm 23.89 \text{ mas yr}^{-1}; \quad (5)$$

$$\kappa_{WE} = -0.1172 \pm 0.0872 \text{ mas yr}^{-1} \text{ Hz}^{-1} = (1.80 \pm 1.34) \times 10^{-17},$$
$$\lambda_{WE} + r_{WE}^{GR} = -37.61 \pm 6.402 \text{ mas yr}^{-1}. \quad (6)$$

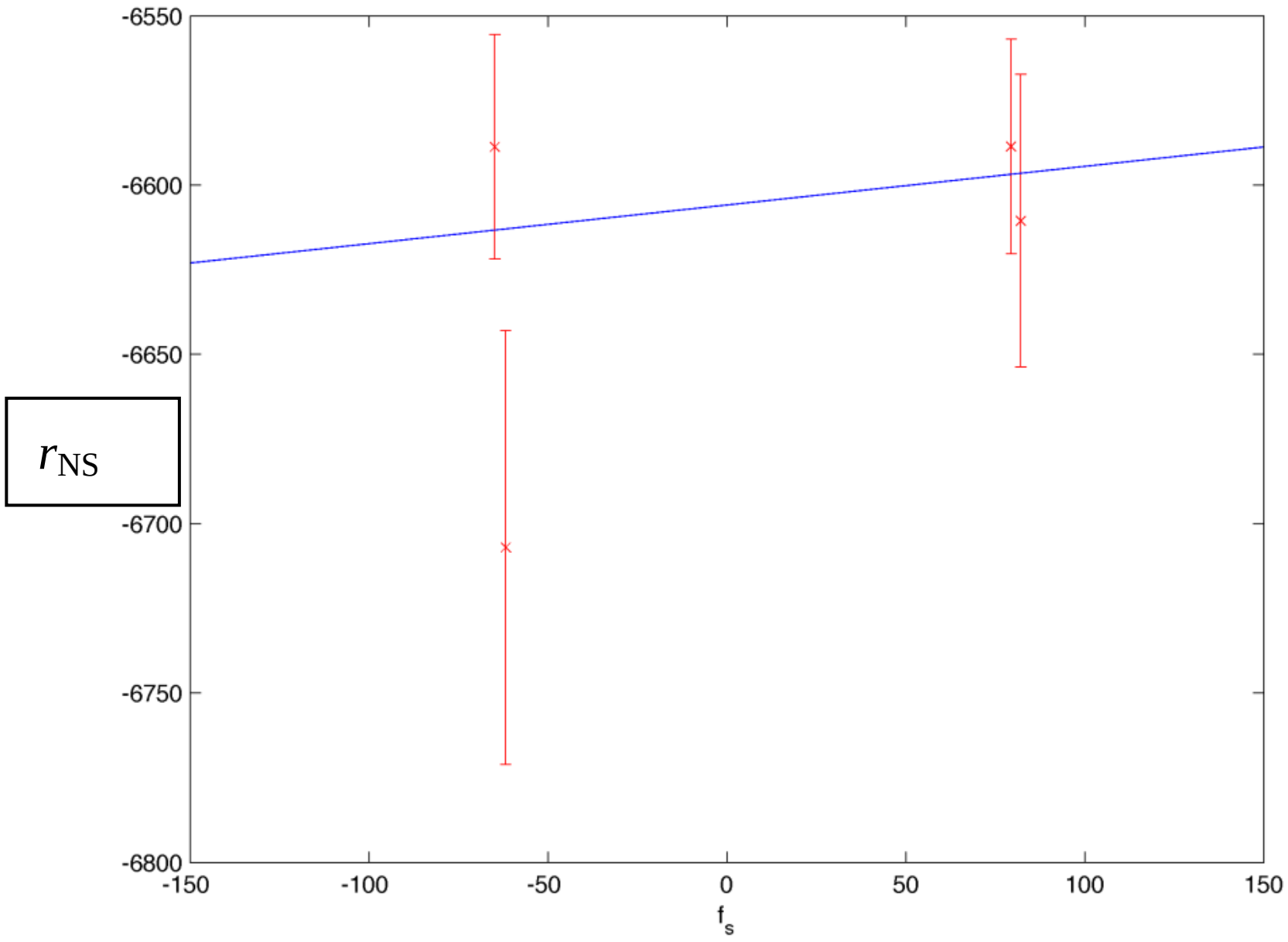

FIG. 1: Fitting $r_{NS}$ to $r_{NS} = r_{NS}^{A} + r_{NS}^{GR} = \kappa_{NS} f_s + \lambda_{NS} + r_{NS}^{GR}$.

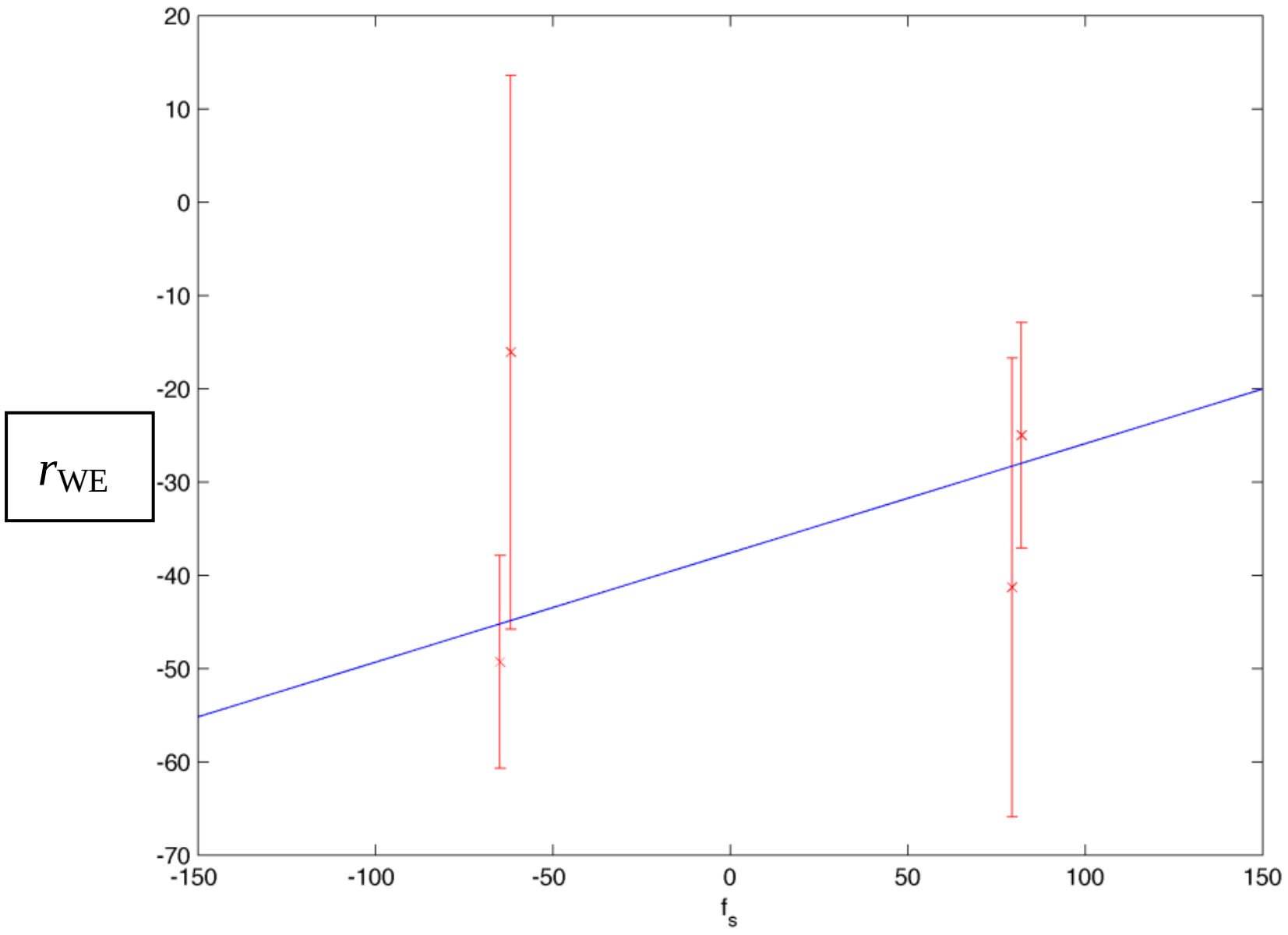

FIG. 2: Fitting $r_{WE}$ to $r_{WE} = r_{WE}^{A} + r_{WE}^{GR} = \kappa_{WE} f_s + \lambda_{WE} + r_{WE}^{GR}$.

From the data fitting, it is not possible to separate $\lambda_{WE}$ with respect to $r_{WE}^{GR}$, and $\lambda_{NS}$ with respect to $r_{NS}^{GR}$. If we treat GP-B experiment as a WEP II experiment, we can subtract the predicted values of general relativity and obtain the WEP II parameter values and constraints. This could be more or less justified, since the general relativistic predictions on LAGEOS were verified to 10-30 % independently [23, 24]. These results are tabulated in the second and third rows of Table III where we list the constraints on λ in all three directions. The 1.34 σ effect of $\kappa_{WE}$ from zero is normal (does not mean a violation) in probability distribution when one has several parameters.

The constraints in the fourth row are obtained as follows. The spin-down rates of four GP-B gyros are accounted for fairly well by standard physics modeling. The unaccounted-for part should not be more than 30 % of the dissipation either way [20]. Therefore the room for anomalous effects should not be more than 30 %. From Table II, 30% of the spin-down rates for G1, G2, G3 and G4 are $5.98 \times 10^{-13}$ $s^{-1}$, $7.01 \times 10^{-13}$ $s^{-1}$, $13.20 \times 10^{-13}$ $s^{-1}$ and $3.60 \times 10^{-13}$ $s^{-1}$. $\lambda_{\text{guide star}}$ should be constrained by all of this. Hence, we list the smallest value in Table III. For the frequency-dependence parameter κ, after all possible combinations are considered, a conservative constraint is listed: within the range of ±80 Hz, the variations are less than $20 \times 10^{-13}$ $s^{-1}$; a conservative estimate would be $< 3 \times 10^{-14}$.

Table III. Test of WEP II regarding to rotational state using rotating quartz balls from GP-B experiment.

| WEP II violation parameter | λ [$s^{-1}$] | κ [dimensionless] |
|---|---|---|
| constraint in the direction (NS) of geodetic effect | $(-0.05 \pm 3.67) \times 10^{-15}$ $s^{-1}$ | $(1.75 \pm 4.96) \times 10^{-17}$ |
| constraint in the direction (WE) of frame dragging effect | $(0.24 \pm 0.98) \times 10^{-15}$ $s^{-1}$ | $(1.80 \pm 1.34) \times 10^{-17}$ |
| constraint in the direction of guide star | $(0 \pm 3.60) \times 10^{-13}$ $s^{-1}$ | $(0 \pm 3) \times 10^{-14}$ |

*Discussions*: (i) GP-B experiment, with its superb accuracy verifies WEP II for unpolarized bodies to an ultimate precision – four-order improvement on the non-influence of rotation on the trajectory, and ultra-precision on the rotational equivalence (no anomalous torques).

(ii) For polarized bodies, the mechanical equivalence of quantum spin and orbital angular momentum is demonstrated to a certain degree [25, 26]. However, there are examples of Lagrange-based theoretical examples of polarized bodies which violate WEP II [5]. These theoretical models may indicate cosmic polarization rotation which are being looked for and tested in the CMB experiments [27]. To look into the future, measurement of gyrogravitational ratio of particle would be a further step [26] towards probing the microscopic origin of gravity. GP-B serves as a starting point for the measurement of gyrogravitational factor of particles.

**Acknowledgements**

The ultra-precision GP-B experiment makes an ultimate test of WEP II for unpolarized bodies possible. I would like to thank Francis Everitt, Sasha Buchman, John Conklin and Mac Keiser and other members of GP-B team for many helpful discussions. This work is supported in part by the National Science Council (Grant Nos NSC97-2112-M-007-002 and NSC98-2112-M-007-009).

**References**

[1] G. Galilei, *Discorsi e dimostriazioni matematiche intorno a due nuove scienze*, (Elzevir, Leiden, 1683); English Translation by H. Crew and A. de Salvio, *Dialogues Concerning Two New Sciences* (Macmillan, New York, 1914) (reprinted by Dover, New York, 1954)

[2] A. Einstein, Über das relativitätsprinzip und die aus demselben gezogenen folgerungen, *Jahrb. Radioakt. Elektronik* **4**, 411 (1907); Corrections by Einstein in *Jahrb. Radioakt. Elektronik* **5**, 98 (1908); English Translations by H. M. Schwartz in *Am. J. Phys.* **45**, 512, 811, 899 (1977).

[3] C. W. Misner, K. S. Thorne and J. A. Wheeler, *Gravitation* (Freeman, San Francisco, 1973)

[4] W.-T. Ni, Weak equivalence principles and gravitational coupling, *Bull. Am. Phys. Soc.* **19**, 655 (1974).

[5] W.-T. Ni, Equivalence principles and electromagnetism, *Phys. Rev. Lett.* **38**, 301, (1977).

[6] H. Hayasaka and S. Takeuchi, Anomalous weight reduction on a gyroscope's right rotation around the vertical axis on the earth, *Phys. Rev. Lett.* **63**, 2701, (1989).

[7] J. E. Faller, W. J. Hollander, P. G. Nelson and M. P. McHugh, Gyroscope-weighing experiment with a null result, *Phys. Rev. Lett.* **64**, 825 (1990)

[8] T. J. Quinn and A. Picard, The Mass of Spinning Rotors - no Dependence on Speed or Sense of Rotation, *Nature* **343**, 732 (1990).

[9] J. M. Nitschke and P. A. Wilmarth, Null Result for the Weight Change of a

Spinning Gyroscope, *Phys. Rev. Lett.* **64**, 2115-2116, 1990

[10] A. Imanishi, K. Maruyama, S. Midorikawa and T. Morimoto, Observation against the weight reduction of spinning gyro, *J. Phys. Soc. Japan* **60**, 1150,1991

[11] J. Luo, Y. X. Nie, Y. Z. Zhang and Z. B. Zhou, Null result for violation of the equivalence principle with free-fall rotating gyroscopes, *Phys. Rev. D* **65,** 042005 (2002).

[12] Z. B. Zhou, J. Luo, Q. Yan, Z. G. Wu, Y. Z. Zhang and Y. X. Nie, New upper limit from terrestrial equivalence principle test for extended rotating bodies, *Phys. Rev. D* **66** 022002, (2002).

[13] C. W. F. Everitt, D. B. DeBra, B. W. Parkinson, J. P. Turneaure, J. W. Conklin, M. I. Heifetz, G. M. Keiser, A. S. Silbergleit, T. Holmes, J. Kolodziejczak, M. Al-Meshari, J. C. Mester, B. Muhlfelder, V. Solomonik, K. Stahl, P. Worden, W. Bencze, S. Buchman, B. Clarke, A. Al-Jadaan, H. Al-Jibreen, J. Li, J. A. Lipa, J. M. Lockhart, B. Al-Suwaidan, M. Taber, S. Wang, Gravity Probe B: Final Results of a Space Experiment to Test General Relativity, *Phys. Rev. Lett.,* in press (2011) ; arXiv:1105.3456.

[14] C. W. F. Everitt *et al.*, Gravity Probe B data analysis --- status and potential for improved accuracy of scientific results, *Space Sci. Rev.*, **148**, 53 (2009).

[15] M. Heifetz *et al.*, The Gravity Probe B data analysis filtering approach, *Space Sci. Rev.*, **148**, 411 (2009).

[16] G. M. Keiser *et al*, Misalignment and resonance torques and their treatment in the GP-B data analysis, *Space Sci. Rev.*, **148**, 383 (2009).

[17] B. Muhlfelder *et al*, GP-B systematic error determination, *Space Sci. Rev.*, **148**, 429 (2009).

[18] A. Silbergleit *et al*, Polholde motion, trapped flux, and the GP-B science data analysis, *Space Sci. Rev.*, **148**, 397 (2009).

[19] J. W. Conklin, Estimation of the mass center and dynamics of a spherical test mass for gravitational reference sensors, Ph. D. thesis, Stanford University (2009).

[20] S. Buchman, private communication, 2009.

[21] G. M. Keiser, private communication, 2011.

[22] Gravity Probe B, Gravity Probe B Quick Facts *Gravity Probe B --- Post Flight Analysis Final Report* (2007).

[23] I. Ciufolini and E. C. Pavlis, Nature 431, 958 (2004).

[24] L. Iorio, *Space Science Reviews* **148**, 363 (2009).

[25] W.-T. Ni, Macroscopic manifestations of the quantum spins: inertial torques on the spin-polarized bodies, *Bull. Am. Phys. Soc.* **29,** 751 (1984).

[26] W.-T. Ni, Searches for the role of spin and polarization in gravity, *Rep. Prog.*

*Phys.* **73**, 056901 (2010) ; and references therein.
[27] W.-T. Ni, From equivalence principles to cosmology: cosmic polarization rotation, CMB observation, neutrino number asymmetry, Lorentz invariance and CPT, *Prog. Theor. Phys. Suppl.* **172**, 49 (2008) [arXiv:0712.4082]; and references therein.